# First Demonstration of Second-order Training of Deep Neural Networks with In-memory Analog Matrix Computing

Saitao Zhang†, Yubiao Luo†, Shiqing Wang, Pushen Zuo, Yongxiang Li, Lunshuai Pan, Zheng Miao, Zhong Sun*
Peking University, Beijing Advanced Innovation Center for Integrated Circuits, Beijing 100871, China.
†Equal Contribution. *Email: zhong.sun@pku.edu.cn

***Abstract*—**Second-order optimization methods, which leverage curvature information, offer faster and more stable convergence than first-order methods such as stochastic gradient descent (SGD) and Adam. However, their practical adoption is hindered by the prohibitively high cost of inverting the second-order information matrix, particularly in large-scale neural network training. Here, we present the first demonstration of a second-order optimizer powered by in-memory analog matrix computing (AMC) using resistive random-access memory (RRAM), which performs matrix inversion (INV) in a single step. We validate the optimizer by training a two-layer convolutional neural network (CNN) for handwritten letter classification, achieving 26% and 61% fewer training epochs than SGD with momentum and Adam, respectively. On a larger task using the same second-order method, our system delivers a 5.88× improvement in throughput and a 6.9× gain in energy efficiency compared to state-of-the-art digital processors. These results demonstrate the feasibility and effectiveness of AMC circuits for second-order neural network training, opening a new path toward energy-efficient AI acceleration.

## I. Introduction

Training and inference are the two fundamental operational phases of neural networks. Training is essential for determining valid model weights, which are then deployed on target devices for inference [1-2]. Due to its more complex computational nature, training poses greater challenges than inference and demands significantly more processing power. Traditional first-order methods utilize only gradient information, without accounting for the curvature of the loss landscape, which can lead to slow convergence and, in some cases, suboptimal performance [3]. In contrast, second-order methods explicitly capture this curvature by performing INV on the second-order information matrix (**Fig. 1a**). This allows them to precondition the gradient through automatic scaling and rotation, resulting in more effective optimization (**Fig. 1b**) [4]. However, implementing second-order optimization on conventional digital computers faces several challenges: (1) the high computational cost of matrix inversion (INV) due to its cubic complexity; (2) the limited memory capacity of digital systems to store the rapidly growing parameters of AI models; and (3) inefficient data transmission in traditional architectures (**Fig. 2**) [5].

Recently, an RRAM-based in-memory AMC circuit was developed to perform real-valued INV in a single step, leveraging the high parallelism and storage density of RRAM [6]. Meanwhile, Kronecker-factored approximate curvature (KFAC) can reduces the size of the matrix to be inverted by approximating the fisher information matrix (FIM), a form of second-order information [7]. In this work, we develop a second-order optimizer based on a scalable, high-precision in-memory AMC system. By approximating FIM using KFAC, we experimentally demonstrate second-order training of a two-layer CNN for handwritten letter classification, achieving performance comparable to the software baseline. Evaluation on a larger training task reveals significant improvements in both throughput and energy efficiency.

## II. Analog Matrix Computing Circuit

**Fig. 3a** illustrates the in-memory AMC circuit based on a 1T1R array, which solves the equation $\boldsymbol{Ax}=\boldsymbol{b}$. The matrix $\boldsymbol{A}^{(m\times m)}$ is mapped as the conductance values of RRAM devices, referenced to a unit conductance value $G_0$. To represent negative values, a column-wise splitting scheme with analog inverters is used, decomposing the original matrix as $\boldsymbol{A}=\boldsymbol{A}_{pos}-\boldsymbol{A}_{neg}$, where both $\boldsymbol{A}_{pos}^{(m\times m)}$ and $\boldsymbol{A}_{neg}^{(m\times m)}$ contains only positive entries [8]. Each source line (SL) in the array connects to the inverting input terminal of an operational amplifier (OA). The bit-lines (BLs) of the $\boldsymbol{A}_{pos}$ array are connected to the OA outputs, while those of the $\boldsymbol{A}_{neg}$ array are connected to the analog inverter outputs. **Fig. 3b** shows an optical photograph of the PCB system implementing the AMC circuit. Input voltages, generated by DACs, are applied to the SLs through unit conductance elements and represent $-\boldsymbol{b}^{(m\times 1)}$. During operation, a bias voltage is applied to the word lines (WLs) by the Arduino MCU to control the transistors, resulting in that output voltages of OAs correspond to the solution vector $\boldsymbol{x}^{(m\times 1)}=\boldsymbol{A}^{-1}\boldsymbol{b}$, which is then captured by ADCs. All DACs and ADCs are controlled by MCU.

## III. Characteristics of RRAM Devices

To construct the closed-loop AMC circuit in **Fig. 3a**, and considering device and circuit reliability, the front-end transistor array was fabricated using 350 nm CMOS technology by a commercial foundry (**Fig. 4a**). The back-end Pt/Ta/$HfO_2$/Pt RRAM was then integrated in a university laboratory (**Fig. 4b**). An optical photograph of the 8×8 1T1R array is also shown in **Fig. 4c**. **Fig. 5** presents the I-V characteristics of a typical 1T1R cell under 8 distinct gate voltage conditions, with 20 cycles per condition. **Fig. 6** shows the retention characteristics of the RRAM for over 100,000 seconds across 8 analog conductance states. **Fig. 7** displays the cumulative distribution function (CDF) test results for the 8 states (ranging from 20 μS to 220 μS). The cells can be programmed to arbitrary conductance states using a write-verify method with a predefined tolerance (**Fig. 8**). **Fig. 9** shows the repeated conductance programming results of 32 1T1R cells used in the real-valued 4×4 INV circuit experiments performed in this work, with a write error of 10 μS. It includes 2,800

RRAM array updates, resulting in 89,600 conductance values in the figure. To evaluate endurance, the two extreme conductance states (20 μS and 200 μS) were selected for testing. **Fig. 10** shows that the RRAM device remains functional after more than 10,000 cycles.

## IV. In-memory Second-order Optimizer Architecture

By approximating the inverse of the layer-wise diagonal-block of the FIM, $\boldsymbol{F}_l^{(f_l \times f_l)}$, as the Kronecker product of the inverse of two smaller matrices $\boldsymbol{A}_{l-1}^{(a_l \times a_l)}$ and $\boldsymbol{G}_l^{(g_l \times g_l)}$, KFAC decomposes the INV of the full FIM $\boldsymbol{F}^{(\Sigma_l f_l \times \Sigma_l f_l)}$ into INVs of multiple small matrices (**Fig. 11a**). Damping factors $\alpha_l$ and $\beta_l$ are also added to regularize the optimization direction. After vectorizing, the update vector $\nabla\boldsymbol{\theta}_l$ for layer $l$ can be derived and used to update the model parameters. In the training process of our second-order optimizer (**Fig. 11b**), forward and backward passes are executed digitally to compute the gradient $\nabla\boldsymbol{w}_l$, Kronecker factors $\boldsymbol{A}_{l-1}$, $\boldsymbol{G}_l$, and damping factors $\alpha_l, \beta_l$. Then, $\nabla\boldsymbol{w}_l$, $\boldsymbol{G}_l$ and $\beta_l$ are transmitted to the multi-stage BlockAMC module, which partitions the large-scale INV operation into smaller ones to compute$(\boldsymbol{G}_l + \beta_l\boldsymbol{I})^{-1}\nabla\boldsymbol{w}_l$. This intermediate result is then passed to a subsequent BlockAMC module, along with $\boldsymbol{A}_{l-1}$ and $\alpha_l$, to compute the final update $(\boldsymbol{G}_l + \beta_l\boldsymbol{I})^{-1}\nabla\boldsymbol{w}_l(\boldsymbol{A}_{l-1} + \alpha_l\boldsymbol{I})^{-1}$, with appropriate matrix transposition. After vectorization, the update vector $\nabla\boldsymbol{\theta}_l$ is returned to the digital system for parameter updates.

**Figs. 11c,d** illustrate details of the specially designed algorithm flow. Multi-stage BlockAMC performs large-scale INV using a recursive algorithm and RRAM arrays sized to meet hardware constraints (**Fig. 11c**) [9]. To ensure precision, a high-precision inversion (HP-INV) algorithm is integrated at the bottom layer of the multi-stage BlockAMC. This is implemented through an iterative process that combines low-precision INV operations (performed by AMC circuits) with high-precision matrix-vector multiplication (MVM) operations (executed by digital systems), as shown in **Fig. 11d** [10].

## V. Experimental results

To validate our optimizer, we designed a customized neural network (**Fig. 12**) for handwritten letter classification, consisting of a convolutional layer, an average pooling layer, and a fully connected (FC) layer. The training set includes four classes of handwritten letters (m, a, n, and c), with 50 images per class randomly selected from the EMNIST dataset [11], totaling 200 images. The test set contains 400 images in total, with 100 images per class. To match the model's scale, the image data were downsampled from 28×28 to 8×8 before being fed into the model. We use mini-batches of size 100, resulting in two update steps per epoch. In the experiment, we train the model for 50 epochs, yielding 100 total update steps. Since the model consists of two layers and each layer requires update vectors for both weights and biases, a total of 400 update vectors are computed using the KFAC-based optimizer.

During training, the multi-stage BlockAMC algorithm employs both uniform and non-uniform partitioning schemes, allowing it to adapt to matrices of varying sizes (**Fig. 13**). The target precision is set to 24-bit fixed-point (equivalent to FP32) for epochs 1-28. To meet higher precision requirement in the later stages, 26-bit fixed-point precision is used for the remaining epochs. We define the relative error as $\|\boldsymbol{v} - \boldsymbol{v}^*\| / \|\boldsymbol{v}^*\|$, where $\boldsymbol{v}^*$ and $\boldsymbol{v}$ represent the ideal and experimental vectors, respectively. The 4×4 AMC circuit performs low-precision INV, yielding 283,606 4×1 vector outputs with a relative error of 58.33% (**Fig. 14**). Through HP-INV iterations, 13,900 high-precision 4×1 vectors are obtained (**Fig. 15**), with an average of 20.4 iteration cycles and a relative error of just 0.013%. These high-precision vectors are then used by the multi-stage BlockAMC to compute 400 update vectors with a final relative error of 4.47% (**Fig. 16**).

We evaluate our experimental results against those of KFAC, Adam, and SGD with momentum (SGD-m) implemented in software. For training speed comparison, we optimize the hyper-parameters to consistently achieve 100% training accuracy (**Fig. 17**). Despite minor errors, our experimental results align well with the KFAC software baseline, achieving 100% accuracy at epoch 37. This outperforms SGD-m (epoch 50) and Adam (epoch 94), requiring 26% and 61% fewer epochs, respectively (**Fig. 18a**). During loss minimization, the second-order training also exhibits smoother convergence and stronger optimization capability compared to first-order methods (**Fig. 18b**). The t-distributed stochastic neighbor embedding (t-SNE) results in **Fig. 19** shows that the trained model achieves effective feature separation, and final trained model yields an 85.1% classification accuracy on the test set (**Fig. 20**).

## Conclusion

In this work, for the first time, we demonstrate a second-order optimizer using RRAM-based in-memory AMC circuit. The optimization performance of our system matches software baseline well while achieving faster training speeds than first-order methods. Based on the performance analysis of training a larger neural network with KFAC [12], **Table. 1** presents benchmark results comparing this work with other state-of-the-art digital processors [13-15], where this work achieves 5.88× higher throughput and 6.9× greater energy efficiency. By bridging the gap between theoretical advancements and practical implementation, this work opens new possibilities for accelerating second-order training in the AI field.

## Acknowledgment

This work has received funding from Beijing Natural Science Foundation (4252016), National Key R&D Program of China (2020YFB2206001), and the 111 Project (B18001).

## Second-order Training

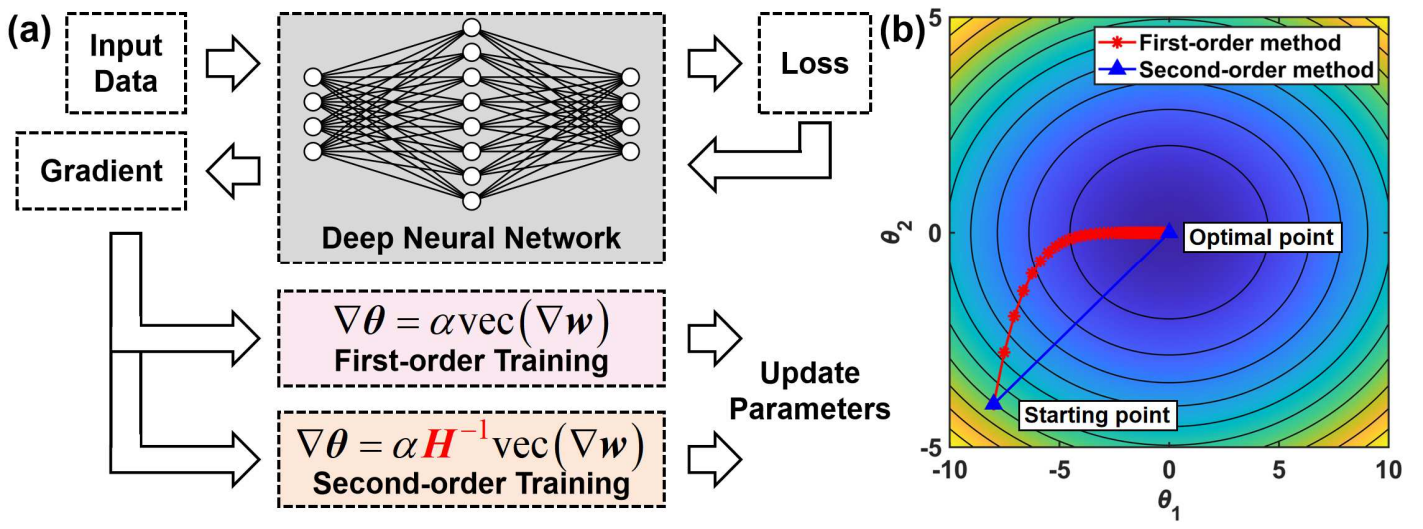

Fig. 1. (a) By inversing the second-order information matrix $\boldsymbol{H}$, second-order training methods precondition the gradient. (b) Optimization trajectory of first-order and second-order method on a simple 2D quadratic function. The second-order method converges faster than the first-order method.

## Motivation

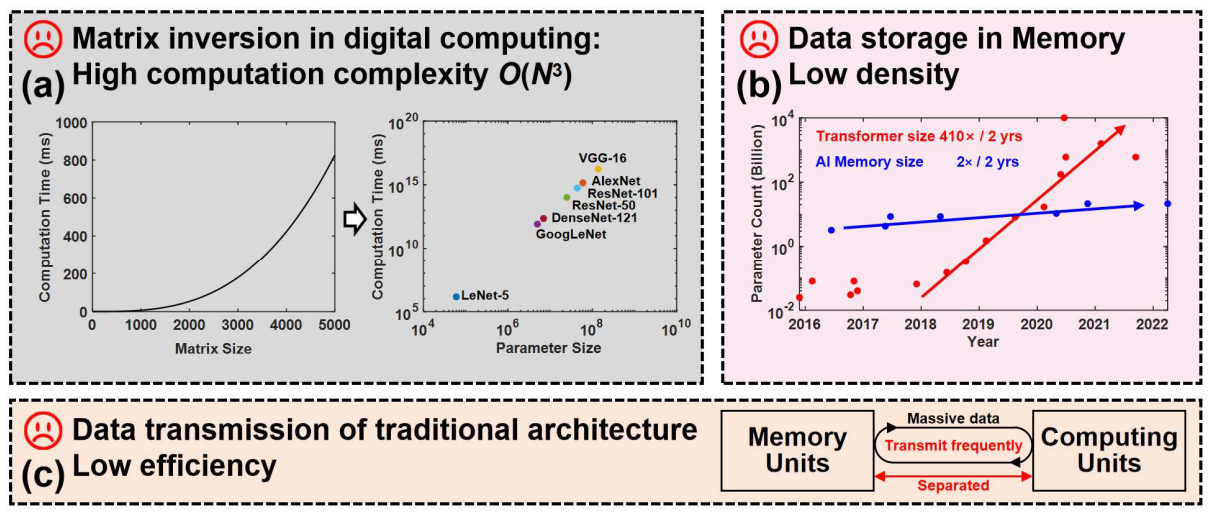

Fig. 2. (a) Matrix inversion are expensive in digital computing. (b) Digital system cannot meet the increasingly growing parameters requirements of AI model. (c) Inefficient Data movement in traditional architecture.

## In-memory Analog Matrix Computing Circuit

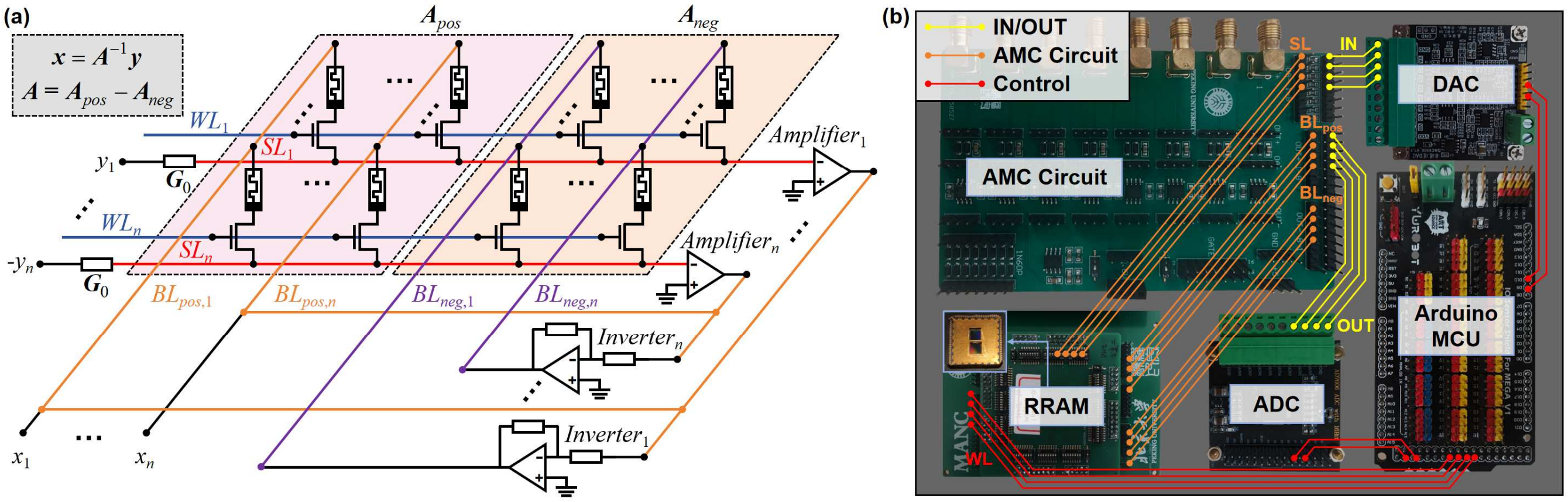

Fig. 3. (a) RRAM-based in-memory AMC circuit for inverting a real-valued matrix in one step. By splitting the matrix $\boldsymbol{A}$ as $\boldsymbol{A}_{pos}$−$\boldsymbol{A}_{neg}$ and incorporating analog inverters, AMC circuit can operate real-valued INV in one step. Each analog inverters consist of an OA and two resistors, with the OA identical to those connected to the rows. (b) Optical photograph of the PCB system implementing the in-memory analog matrix computing circuit.

## Characterization of 1T1R Array

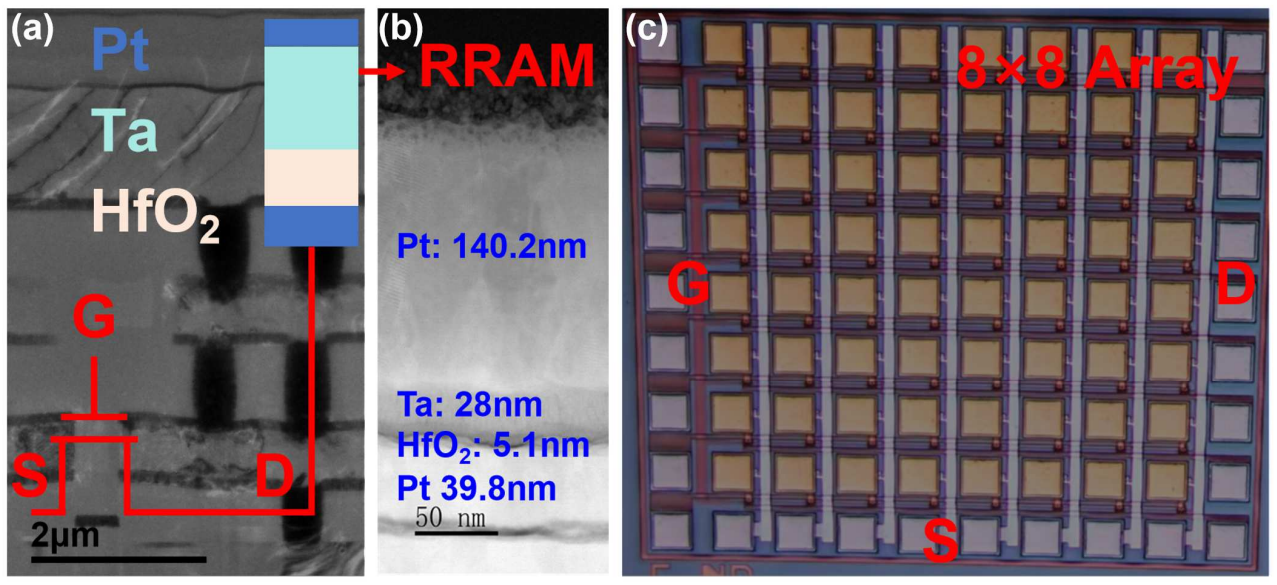

Fig. 4. TEM images of (a) the front-end transistor, and (b) the back-end Pt/Ta/$HfO_2$/Pt RRAM, which is integrated at the drain of the corresponding transistor. (c) Optical microscope image of an 8×8 RRAM array.

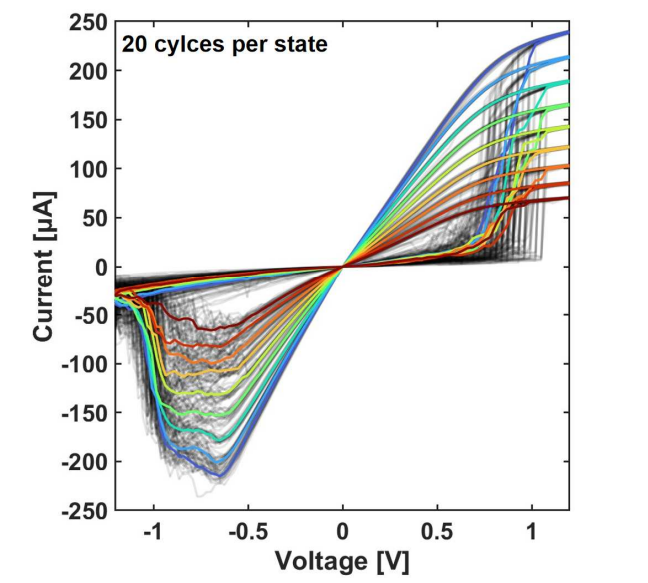

Fig. 5. I-V characteristics of a typical RRAM device under 8 different gate voltage, with 20 cycles for each case.

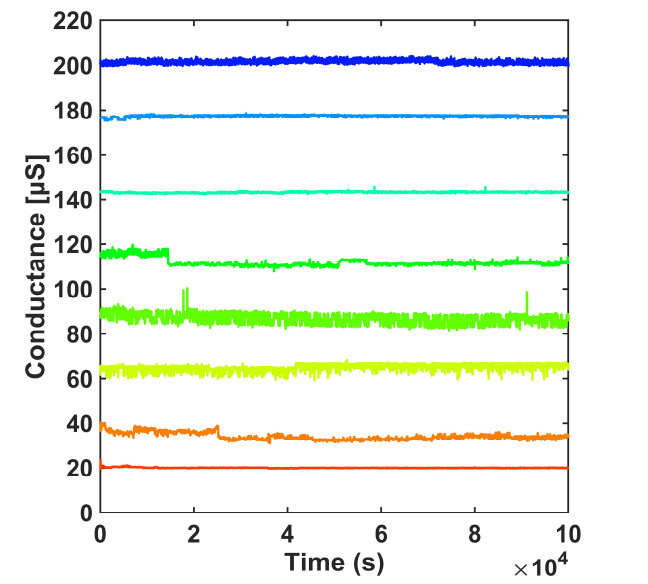

Fig. 6. Retention characteristic of the RRAM device with 8 conductance states (>100,000s).

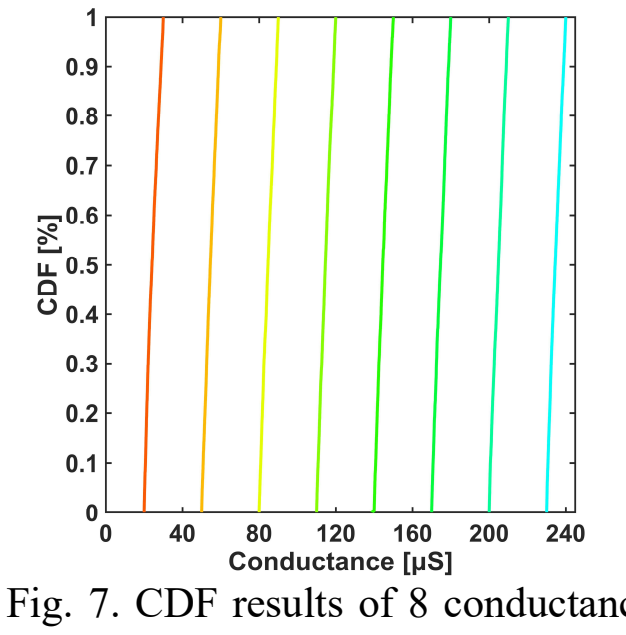

Fig. 7. CDF results of 8 conductance states programmed using the write-verify method.

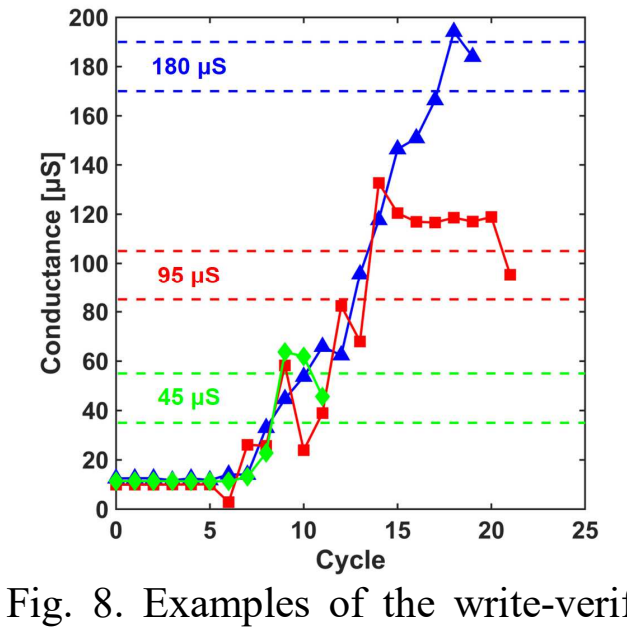

Fig. 8. Examples of the write-verify method for programming RRAM devices.

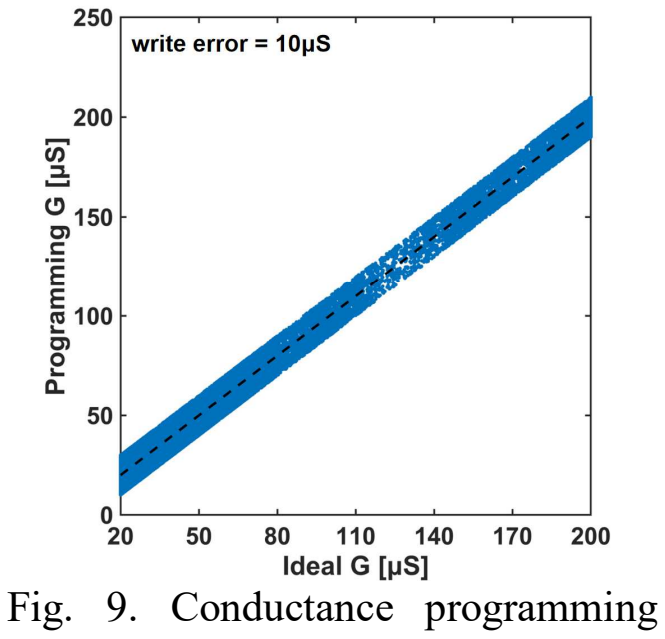

Fig. 9. Conductance programming results of RRAM ranging from 20 to 200 μS, with a write error of 10 μS.

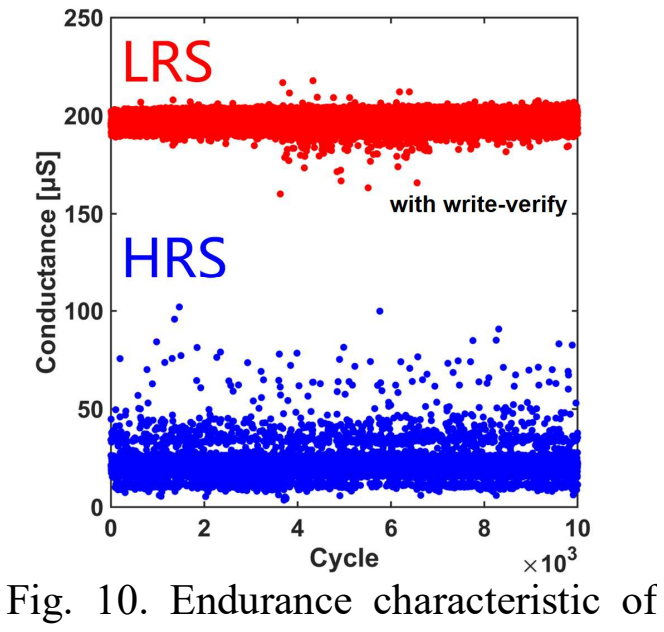

Fig. 10. Endurance characteristic of the RRAM device (>10,000 cycles).

## In-memory Second-order Optimizer

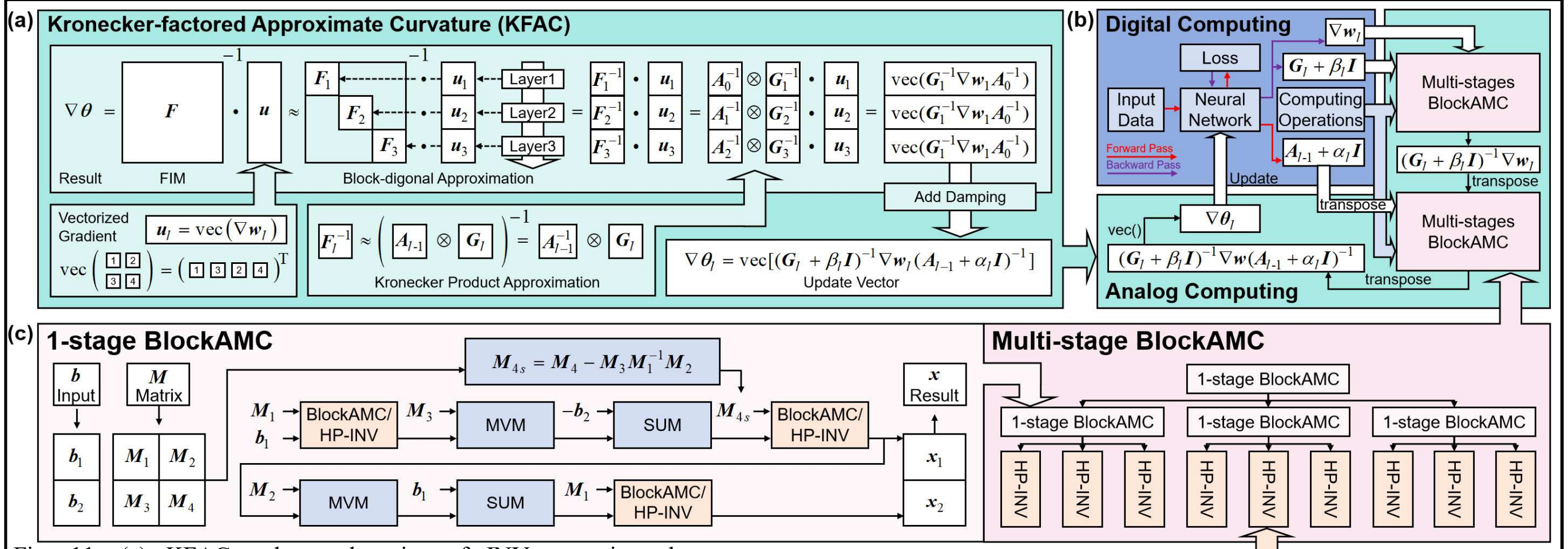

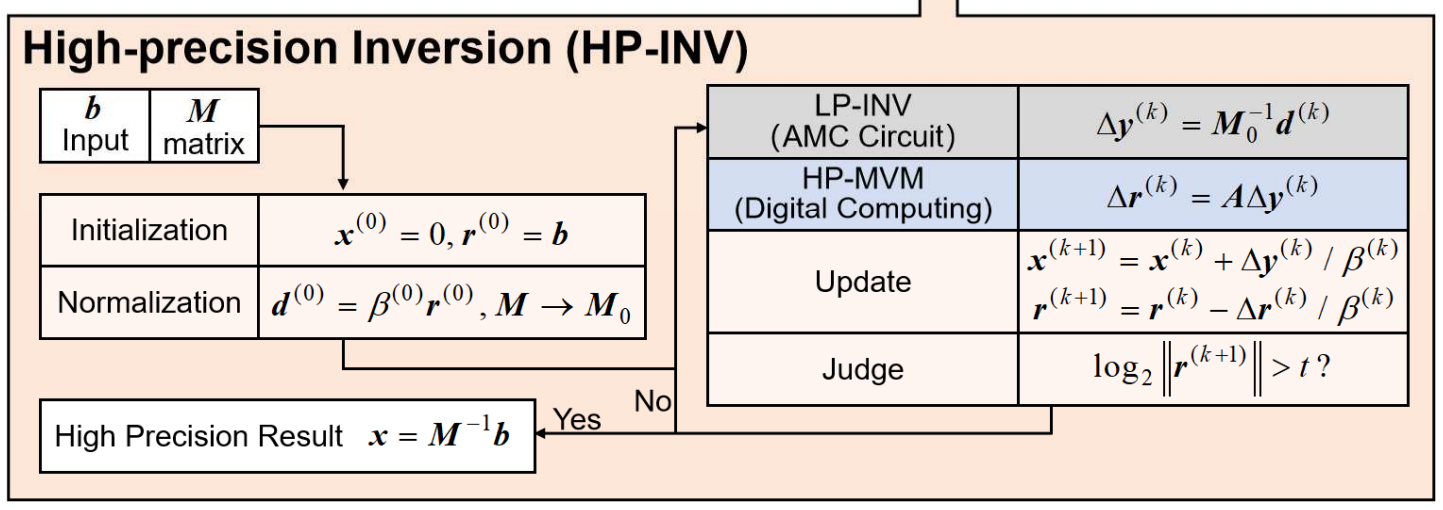

Fig. 11. (a) KFAC reduces the size of INV operations by approximating the Fisher information matrix as a layer-wise block-diagonal structure with Kronecker products. The update vector is obtained by vectorizing the results. (b) KFAC-based second-order optimizer implemented with in-memory AMC. Two serial multi-stage BlockAMC modules perform the inversions of the two Kronecker factors, respectively. (c) BlockAMC recursively partitions a large matrix into smaller submatrices that fit within the size constraints of RRAM arrays. (d) The HP-INV algorithm combines low-precision INV using AMC circuits with high-precision MVM via digital computing to achieve high-precision INV.

## Experimental Demonstration of Second-order Training

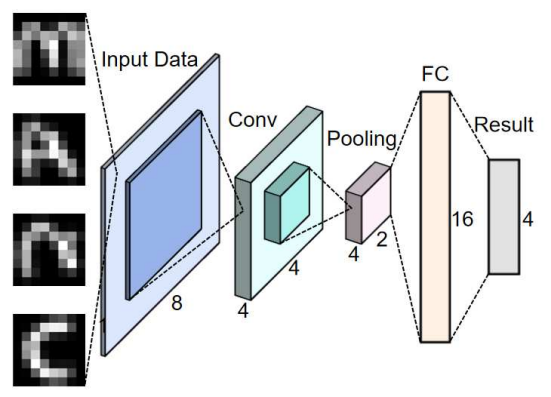

Fig. 12. Structure of customized neural network. Four classes of handwritten letters make up the dataset.

| Layer | BlockAMC | | | 4-INV Number per BlockAMC | BlockAMC Number |
|---|---|---|---|---|---|
| Conv Layer | 25 | 9 | 8 | 3 | 4 |
| | | 16 | | 3×3 | 4 |
| | | 9 | 8 | 3 | 4 |
| | 4 | | | 1 | 25 |
| | 4 | | | 1 | 1 |
| FC Layer | 16 | | | 3×3 | 4 |
| | 4 | | | 1 | 16 |
| | 4 | | | 1 | 1 |
| Total 4-INV Number | | | | 139 | |

Smallest matrix size is 4

Fig. 13. Matrix partitioning during the computation of update vectors for two layers using the BlockAMC algorithm.

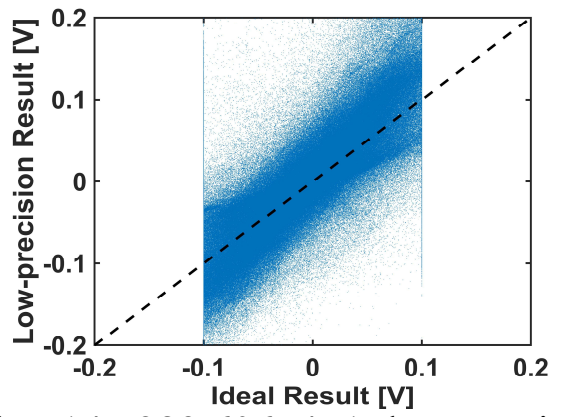

Fig. 14. 283,606 4×1 low-precision results obtained from the AMC circuit throughout the entire training process, showing a relative error of 58.33%.

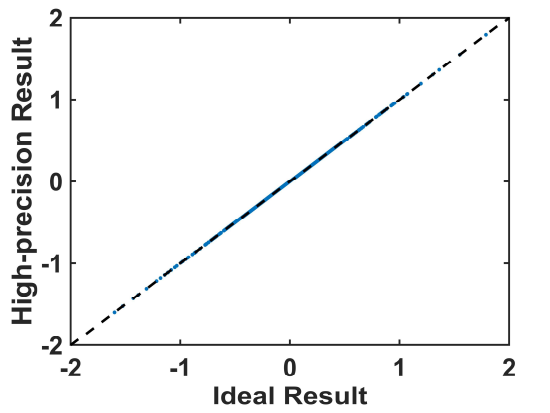

Fig. 15. 13900 4×1 obtained from the HP-INV algorithm, based on 283,606 low-precision results, with a relative error of 0.013%.

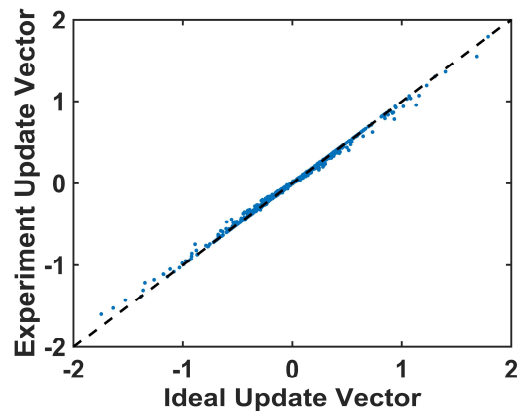

Fig. 16. 400 update vector results from BlockAMC algorithm based on 13,900 high-precision results, with a relative error of 4.47%.

| General Hyper-parameters | | | |
|---|---|---|---|
| Hyper-parameters | KFAC | SGD-m | Adam |
| Learning Rate | 0.3 | 1 | 0.1 |
| Weight Decay | 0.00001 | 0.0003 | 0.003 |
| Exponential LR Rate | 1 | 0.96 | 0.96 |
| Specific Hyper-parameters | | | |
| For KFAC, Inverse frequency T=1, damping=0.03 | | | |
| For Adam, $\beta_1=\beta_2=0.9$ | | | |
| For SGD-m, $\beta=0.9$, using nesterov momentum | | | |

Fig. 17. Hyper-parameters used for KFAC, SGD-m and Adam in the training experiments and baseline comparisons.

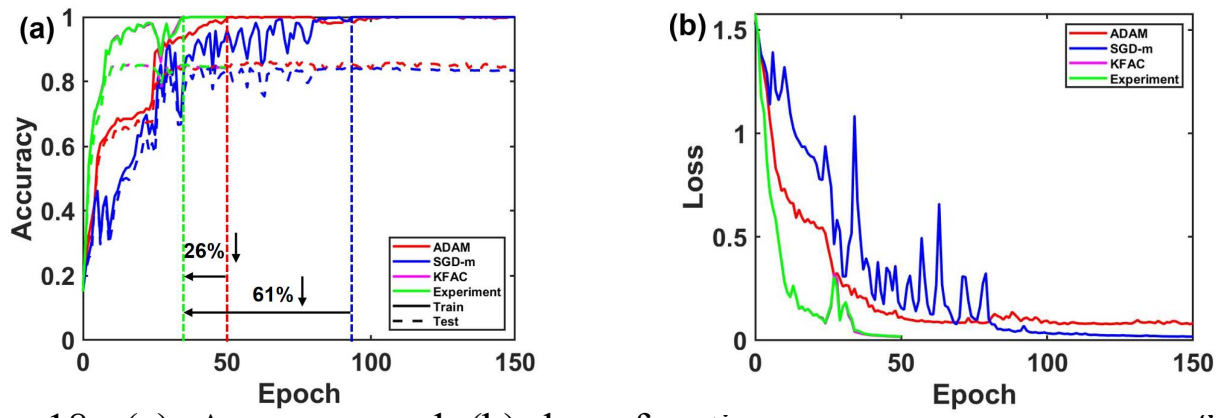

Fig. 18. (a) Accuracy and (b) loss function convergence curves of the customized neural network trained with ADAM, SGD-m, KFAC, and experiment results. KFAC demonstrates the fastest convergence, and the experimental results closely match the ideal outcomes.

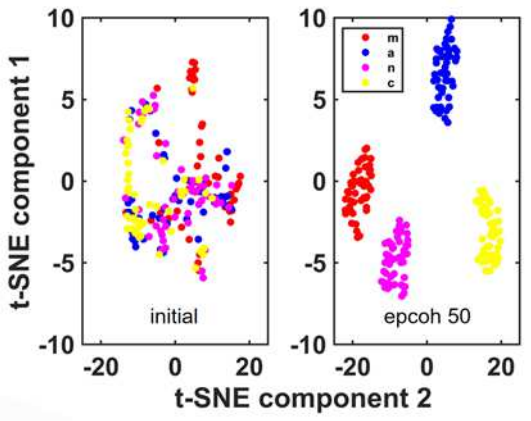

Fig. 19. t-SNE results of FC layer in epoch 1 and epoch 50.

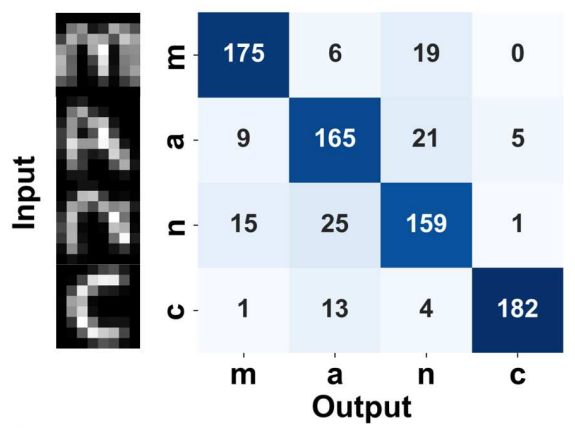

Fig. 20. Confusion matrix of the classification results at epoch 50, showing 85.1% accuracy.

| Target Probelm | Dataset | MNIST | | |
|---|---|---|---|---|
| | Optimizer | KFAC | | |
| | model | Autoencoder [12] | | |
| Approach | ASIC [13] | Nvidia H100 core [14] | AMD Vega 20 core [15] | This Work |
| Hardware | 28nm CMOS | 4nm CMOS | 7nm CMOS | RRAM |
| In-Memory | No | No | No | Yes |
| Anlog or digital | Digital | Digital | Digital | Analog+Digital |
| Precision | FP 32 | FP 32 | FP 32 | FP 32 (analog) |
| Throughput [OPS] | $1.85\times10^{9}$ | $3.21\times10^{9}$ | $2\times10^{9}$ | $1.89\times10^{10}$ (5.8×) |
| Energy Efficiency [OPS/W] | $1.03\times10^{11}$ | $6.7\times10^{10}$ | $4.92\times10^{10}$ | $7.18\times10^{11}$ (6.9×) |

Table. I. Benchmarking results of energy efficiency of AMC and state-of-the-art digital processors in training a large neural network [12] using KFAC, including one ASIC [13] and two GPUs [14,15].